\documentstyle[preprint,prc,aps]{revtex}
\begin{document}
\draft

\title{ The State Dependence Calculation of Three-body \\Cluster Energy
for Nuclear  Matter}
\vskip 10truecm
\author{{ H. R. Moshfegh$^\ast$ and M. Modarres$^{\ast\dagger}$}\\
Physics Department, Tehran University
North-Karegar Ave., Tehran, { Iran}$^\ast$\\
and\\
Center for Theoretical Physics
and Mathematics
AEOI P.O. Box 11365-8486 Tehran, { Iran}$^\dagger$}
\maketitle
\begin{abstract}
It is shown that the method of lowest order constrained variational (LOCV)
which is based on the cluster expansion theory is a reliable many-body
technique to calculate the nuclear matter equation of state. In this
respect, the state dependent  correlation functions and the effective
interactions which have been produced  by the LOCV calculation with the Reid
soft core and the $\Delta$-Reid  interactions are used to estimate the size
of higher order cluster terms  such as the effect of three-body cluster
energy on nuclear matter. We  find that the three-body cluster  energy is
less than 1 MeV beyond the nuclear matter saturation density  and it has
weaker density dependence than our previous  calculation with the
state-averaged correlation function and the effective  interaction. Finally
we conclude that the LOCV method is good enough to  calculate other
properties of quantal fluids.
\end{abstract}

\newpage

\section{Introduction}

The method of lowest constrained variational (LOCV) was developed in
1977-1979 [1,2] to calculate the bulk properties of homogeneous nuclear
fluid such as the saturation energy and density, the surface energy and the
asymmetrical coefficient in the semi-empirical formula, by using the
realistic nucleon-nucleon interactions i.e. the Reid [3] and the $\Delta $%
-Reid [2] potentials. The LOCV method was successfully extended further for
finite temperature calculation [4] and various thermodynamics properties of
nuclear matter were evaluated. Recently this approach was reformulated to
include more sophisticated interactions such as the $V_{14}$ [5], the $%
AV_{14}$ and the new Argonne $AV_{18}$ [6] potentials. In these calculations
[5,6] it was shown that the new $AV_{18}$ potential, like the other N-N
interactions, dose over bind nuclear matter and a very good agreement was
found between the LOCV results and more complicated approaches such as the
Bruckner-Hartree- Fock (BHF), the Correlated Basis Function (CBF), the
Bruckner-Bethe (BB) and the Variational Hypernetted Chain (VHC) techniques.
The results of these calculations are demonstrated in table 1 by presenting
the saturation energy, the saturation density and the incompressibility of
nuclear matter for different potentials and different many-body methods from
reference [6]. According to this table, it is seen that only the results of
the $AV_{14}$ and the $\Delta $-Reid potentials are close to the empirical
values and on average the inclusion of density dependence three-nucleon
interaction (TNI) improves the calculations.

In order to test the convergence of our LOCV results for nuclear matter,  we
performed calculations beyond the lowest order (the two-body cluster  term)
and the three-body cluster energy was calculated with the state-averaged
correlation function [10] which in turn was generated by using the
state-dependent  LOCV correlation functions. The smallness of the
normalization parameter  (the convergence parameter) and the three-body
cluster energy where indicated that  at least up to the twice nuclear matter
density, our expansion converges  reasonably and it is a good approximation
to stop after the two-body  cluster terms [2,5,10].
LOCV formalism has several advantages with respect to the other  many-body
techniques like the Brueckner-Bethe (BB) [7], the Variational Hypernetted
Chain (VHC) [8], the Correlated  Basis Function (CBF) [8] and the
Brueckner-Hartree-Fock (BHF) [9] which go beyond the lowest order:

1) There is no free parameter in the LOCV method beside the N-N potentials
i.e. it is fully self-consistent.

2) It considers constraint in the form normalization condition [11] to keep
the higher cluster terms as small as possible. As we pointed above, this has
been tested by calculating the three-body cluster terms with the
state-averaged  correlation functions.

3) It assumes a particular form for the long-range part of the  correlation
functions in order to perform an exact functional minimization  of the
two-body energy with respect to the short-range behavior of the  correlation
functions.  4) It dose functional minimization, rather than simply
parameterizing the  short-range behavior of the correlation functions. So in
this respect it  also saves an enormous computational time.

The aim of present work is to calculate the three-body cluster terms with
the state-dependent correlations i.e. without making the state-averaged
approximation. In this work we use the Reid soft core [3] and the $\Delta $%
-Reid [2] potentials, since we are only interested in the size of three-body
cluster energy. Beside this the results of our previous calculation shows
that the $\Delta $-Reid interaction can reasonably reproduce the properties
of nuclear matter. On the other hand the $\Delta $ state is the most
important configuration which modifies the nuclear forces and one may
consider it as the   origin of understanding of the three-body forces [12].

So the paper is organized as follows. We begin section II by describing
briefly the LOCV formalism. Section III-A is devoted to the three-body
cluster  energy and the definition of the state-averaged correlation
function and  the effective interaction. In section III-B we present the
explicit equations  for the three-body cluster energy in terms of the
state-dependent correlation  functions and the effective interactions.
Finally, in the last section  we discuss our results, the  various aspects
of LOCV formalism, the three-body cluster energy and the  uncertainties
involved in the treatment of both the tensor and the channel  dependence of
the realistic interaction forces. In this respect  we explain how it is
possible to believe that the LOCV  method is capable of giving good
agreement with the results obtained by the  more complicated schemes in
which the many-body contributions have been  taken in to the account.

\section{The LOCV formalism}

We consider a trial many-body wave function of the following form to
evaluate the Rayleigh-Ritz upper bound to the ground-state energy:
\begin{equation}
\psi_v=F\Phi
\end{equation}
where $\Phi$ is a slater determinant of the plane waves of A independent
nucleons,
\begin{equation}
\Phi={\cal A}\prod_i \exp(i\vec k_i\cdot \vec r_i)
\end{equation}
$F$ is a A-body correlation operators which will be replaced by a Jastrow
type i.e.
\begin{equation}
F(1 \cdot\cdot\cdot A)={\cal S}\prod_{i>j}f(ij)
\end{equation}
and ${\cal A}$ and ${\cal S}$ are an anti-symmetrizing and a symmetrizing
operators respectively. The cluster expansion energy is written as [13]:
\begin{equation}
E([f])={\frac{{1}}{{A}}} {\frac{{<\psi_v\mid H\mid\psi_v>}}{{%
<\psi_v\mid\psi_v>}}}=  E_1+E_2+E_3+\cdot\cdot\cdot\geq E_0
\end{equation}
where $E_0$ is the through ground-state energy. In the lowest order, we
truncate above series after $E_2$.

The one body term $E_1$ is just the familiar Fermi gas kinetic energy i.e.
\begin{equation}
E_1={\frac{3}{5}}{\frac{{\hbar^2{k_F}^2}}{{2m}}}
\end{equation}
where $k_F=({\frac{3}{2}}\pi^2\rho)^{1/3}$ and $\rho$, are the Fermi
momentum and the density of uniform nuclear matter, respectively.
The two-body energy $E_2$ is:
\begin{equation}
E_2={\frac{1}{{2A}}}\sum_{ij}<ij \mid {\cal V}(12) \mid ij>_a
\end{equation}
and the "effective interaction operator" ${\cal V}(12)$ is given by the
following equation:
\begin{equation}
{\cal V}(12)=-{\frac{{\hbar^2}}{{2m}}}[F(12),[%
\nabla^2_{12},F(12)]]+F(12)V(12)F(12)
\end{equation}
$<ij \mid O \mid ij>_a$ represent the antisymmetrized two-body matrix
element taken with respect to the single-particle plane waves.

The two-body correlation operator $F(12)$ is defined as follows:
\begin{equation}
F(ij)=\sum^4_{\alpha,p=1}f^{(p)}_\alpha O^{(p)}_\alpha(ij)
\end{equation}
$\alpha=\{J,L,S,T\} $ and operators $O^{(p)}_\alpha(ij)$ are written as:
\begin{equation}
O^{p=1,4}_\alpha=1,({\frac{2}{3}}+{\frac{1}{6}}S^I_{12}),({\frac{1}{3}}-{
\frac{1}{6}}S^I_{12}) , S^{II}_{12}
\end{equation}
where $S^{I}_{12}$ is the familiar tensor operator and $S^{II}_{12}$ is its
analogous for the mixed $N$-$\Delta$ channel (in case of the $\Delta$-Reid
potential). The values of $p$ is set to unity for $L=0$ and the spin-
triplet channels with $L\neq J\neq 1$. But for $L=J\pm 1 $ it takes values
of 2 and 3. Finally we have $L=0$ channels which couple the $^1S_0$ channel
to the $^5D_0$ channel (for the $\Delta$-Reid potential) where we set $p=1$
and 4. As in our previous works we also require that the correlation
functions $f^{(1)}_\alpha$,$f^{(2)}_\alpha$ and $f^{(3)}_\alpha$ ($%
f^{(4)}_\alpha$)  heal to the pauli function $f_P(r)$ (zero),
\begin{equation}
f_P(r)=[1-[{\frac{3}{2}}l(k_Fr)]^2]^{-1/2}
\end{equation}
where ($J_J(x)$ are the spherical bessel functions)
\begin{equation}
l(x)={\frac{J_1(x)}{x}}
\end{equation}

The two-body nucleon-nucleon interaction $V(12)$ has the following form,
\begin{equation}
V(12)=\sum V^{(p)}(r_{12}) O^{(p)}_{12}
\end{equation}
and they are taken from references [2,3].

The normalization constraint, $<\psi_v \mid \psi_v>=1$, that we impose on
the channel two-body correlation functions $f^{(p)}_\alpha$ as well as the
coupled and uncoupled differential equations, coming from Euler-Lagrange
equations are described in references [1,2].

\section{The three-body cluster energy}

\subsection{The state-averaged calculation}

In general the three-body cluster term expression in the energy expectation
value has the following form [13],
\begin{equation}
E_3=E_3^{(2)}+E_3^{(3)}
\end{equation}
where
\begin{equation}
E_3^{(2)}=-{\frac{1}{{A}}}\sum_{ijk}h_{ik} W_{ij}
\end{equation}
\begin{equation}
E_3^{(3)}={\frac{1}{{A3!}}}\sum_{ijk} W{ijk} \hspace{.25cm}
\end{equation}
and $h_{ik}$ , $W_{ij}$ and $W_{ijk}$ are defined as,
\begin{eqnarray}
h_{ik}=<ik \mid h(12) \mid ik>_a \hspace{0.97cm}\nonumber\\
W_{ij}=<ij \mid {\cal{V}}(12) \mid ij>_a \hspace{0.9cm}\\
W_{ijk}=<ijk \mid{\cal {V}}(123) \mid ijk>_a\nonumber
\end{eqnarray}
The two-body effective operator ${\cal {V}}(12)$ is given in equation (7)
and $(t(i)=-{\frac{{\hbar^2}}{{2m}}}\nabla_i)$
\begin{eqnarray}
h(12)=F^{\dagger}(12)F(12)-1\hspace{7.8cm} \\
{\cal {V}}(123)={1\over 2}F_3^{\dagger}(123)\,[t(1)+t(2)+t(3)\ ,\ F_3(123)]
\ + adj.\hspace{2.3cm}\nonumber \\
\hspace{2.2cm}+(F_3^{\dagger}(123) V(12) F_3(123)- {\cal {V}}(12)
+same\ for\ pairs\ 23\ and\ 13)
\end{eqnarray}
Now by imposing the Jastrow approximation and ignoring the state and
operator dependence of correlation functions, one can write the above
three-body cluster energy $\bar{E}_3$ expressions in the following form :
\begin{equation}
\bar{E}_3=\bar{E}_{3h}^{(2)}+\bar{E}_{3h}^{(3)}+\bar{E}_{3hh}^{(3)}+\bar{T}%
_{3hh}^{(3)}
\end{equation}
where
\begin{equation}
\bar{E}_{3h}^{(2)}=-{\frac{1}{{A}}}\sum_{ijk}<ik \mid h(r_{13})\mid ik>_a<ij
\mid {\cal {V}}(r_{12}) \mid ij>_a
\end{equation}
\begin{equation}
\bar{E}_{3h}^{(3)}=-{\frac{1}{{A}}}\sum_{ijk}<ijk \mid h(r_{13}){\cal {V}}%
(r_{12})\mid ijk>_a
\end{equation}
\begin{equation}
\bar{E}_{3hh}^{(3)}={\frac{1}{{2A}}}\sum_{ijk}<ijk\mid h(r_{13}) {\cal {V}}%
(r_{12}) h(r_{23})\mid ijk>_a
\end{equation}
\begin{equation}
\bar{T}_{3hh}^{(3)}={\frac{1}{{2A}}}\sum_{ijk}<ijk\mid{\frac{{\hbar^2}}{{2m}}%
} f^2(r_{12})\nabla_2 h(r_{12}) \cdot \nabla_2h(r_{23}) \mid ijk>_a
\end{equation}
and the hole function $h(r_{ij})$ is defined as,
\begin{equation}
h(r_{ij})=f^2(r_{ij})-1
\end{equation}
$<ijk \mid O \mid ijk>_a$ represent the antisymmetrized three-body matrix
element taken with respect to the single-particle plane waves.

Now, we define a state-averaged two-body correlation function [10] as
\begin{equation}
\bar {f^2}(x)={\frac{{\sum_{\alpha,i}(2T+1)(2J+1){\frac{1}{{2}}}%
[1-(-1)^{L+T+S}]f^{(i)^2}_\alpha(x) a^{(i)^2}_\alpha (x)}}{{%
\sum_{\alpha,i}(2T+1)(2J+1){\frac{1}{{2}}}[1-(-1)^{L+T+S}]
a^{(i)^2}_\alpha(x)}}}
\end{equation}
and a state-averaged two-body effective interaction as
\begin{equation}
\bar {{\cal V}}(12)={\frac{{\sum_{\alpha,i}(2T+1)(2J+1){\frac{1}{{2}}}%
[1-(-1)^{L+T+S}]{\cal V}^{(i)}_\alpha(12) a^{(i)^2}_\alpha (x)}}{{%
\sum_{\alpha,i}(2T+1)(2J+1){\frac{1}{{2}}}[1-(-1)^{L+T+S}]
a^{(i)^2}_\alpha(x)}}}
\end{equation}
where the $a^i_{\alpha}(x)$, etc. are
\begin{eqnarray}
&a^{(1)^2}_\alpha(x)=x^2I_L(x)\hspace{5.785cm}\nonumber\\
&a^{(2)^2}_\alpha(x)=x^2(2J+1)^{-1}[(J+1)I_{J-1}(x)+JI_{J+1}(x)]\\
&a^{(3)^2}_\alpha(x)=x^2(2J+1)^{-1}[JI_{J-1}(x)+(J+1)I_{J+1}(x)]\nonumber\\
&a^{(4)^2}_\alpha(x)=x^2I_J(x)\hspace{5.785cm}\nonumber
\end{eqnarray}
with 
\begin{equation}
I_L(x)=48\int^1_0 dzz^2(1-{\frac{3}{2}}z+{\frac{1}{2}}z^3)J^2_L(zx)
\end{equation}
and
\begin{equation}
\bar h(ij)=\bar{f^2}(ij)-1
\end{equation}
Finally after doing some algebra, we get the following three-body cluster
terms in terms $\bar h(ij)$ and $\bar{{\cal V}}(ij)$ as,
\begin{eqnarray}
\bar E_{3h}^{(2)}={{\rho^3}\over {4A}}\int \, d^3r_1d^3r_2d^3r_3 \ \bar h(r_{13}) \,
\bar{\cal V}(r_{12})\hspace{8cm}\nonumber \\
\{4+{1\over 4}l(k_Fr_{23})l(k_Fr_{12})l(k_Fr_{13})-(l^2(k_Fr_{12})
+l^2(k_Fr_{13}))\}\hspace{4cm} \\
\bar E_{3h}^{(3)}=\frac {\rho^3}{4A}\int \, d^3r_1d^3r_2d^3r_3 \ \bar h(r_{13})
\, \bar{\cal V}(r_{12})\hspace{8cm}\nonumber \\
\{ 4+{2 \over {4}}l(k_Fr_{12})l(k_Fr_{23})
l(k_Fr_{13})-l^2(k_Fr_{12})-2l^2(k_Fr_{23}) \}\hspace{4cm} \\
\bar E_{3hh}^{(3)}=\frac {\rho^3}{8A}\int \, d^3r_1d^3r_2d^3r_3 \ \bar h(r_{13})\,
\bar h(r_{23})
\, \bar{\cal V}(r_{12})\hspace{7cm}\nonumber \\
\{ 4+{2 \over {4}}l(k_Fr_{12})l(k_Fr_{23})
l(k_Fr_{13})-l^2(k_Fr_{12})-2l^2(k_Fr_{23}) \} \hspace {4cm}\\
\bar T_{3hh}^{(3)}=\frac {\rho^3}{8A}\int \, d^3r_1d^3r_2d^3r_3
 ({{\hbar^2} \over {4m}}
\bar f^2(r_{13}) \nabla_2 \bar h(r_{12}) \cdot \nabla_2
\bar h(r_{23}))\hspace{4.25cm}\nonumber\\
\{ 4+{2 \over {4}}l(k_Fr_{12})l(k_Fr_{23})
l(k_Fr_{13})-l^2(k_Fr_{12})-2l^2(k_Fr_{23}) \} \hspace{4cm}
\end{eqnarray}
Then we can simply add the contribution of the three-body cluster energy, $%
\bar E_3$, to the two-body energy $E_2$.

In the figure 1 and 2 we plot the contributions of various quantities such
as $\bar{E}_3$, $\bar{E}_{3h}^{(2)}$, $\bar{E}_{3h}^{(3)}$, $\bar{E}_{3h} =
\bar{E}_{3h}^{(2)}+\bar{E}_{3h}^{(3)} $, $\bar{E}_{3hh}^{(3)}$  , $\bar{T}%
_{3hh}^{(3)}$ and $\bar{E}_{3hh}^{(3)}+\bar{T}_{3hh}^{(3)}$ against  density
in nuclear matter for the Reid and the $\Delta$-Reid potentials
respectively. As we expect, it is seen that the main contribution comes from
${\bar E}_{3h}$. Since the other terms namely $\bar{E}_{3hh}$ and $\bar{T}%
_{3hh}$  are smaller than this term, i.e $\bar{E}_{3h} $ with order of $\bar{%
h}(ij)$. It is also seen that  the density dependence of $\bar{E}_{3}$ is
mainly due to ${\bar E}_{3h}$.  On the other hand the size of $\bar{E}_{3hh}$
and $\bar{T}_{3hh}$  are roughly the same but in opposite sign beyond the
nuclear matter density.  So as we mention before, we can conclude that only $
\bar{E}_{3h}$ is important in the three-body cluster energy.

\subsection{The state-dependent calculation}

Regarding our discussion in the section III-A , the main contribution of
three-body cluster energy should come from $E_{3h}^{(2)}$ and $E_{3h}^{(3)}$%
. So in this section we explain how it is possible to write $E_{3h}^{(2)}$
and $E_{3h}^{(3)}$ in terms of $f_\alpha^{(i)}(r)$ and ${\cal V}%
^{(i)}_\alpha(r)$ and calculate them exactly i.e. without making
state-averaged approximation.

By using the addition of angular momentum theorem [14], the partial wave
expansion and various orthogonality relations, the explicit form of $%
E_{3h}^{(2)}$ can be written as following,
\begin{eqnarray}
E_{3h}^{(2)}={-1\over A}\sum_{\{k_ik_jk_k\}}\sum_{LSJT}
\sum_{L^\prime J^\prime S^\prime T^\prime}
\int_0^\infty\int_0^\infty (xy)^2 \, dxdy \,
((-1)^{L+S+T}-1)((-1)^{L^\prime+S^\prime+T^\prime}-1)
\nonumber\\
({{(4\pi)^4}\over{\Omega^2}})h^{L^\prime S^\prime J^\prime T^\prime}(x){\cal V}^{LSJT}(y)
J_L^2(k_{ij}x)J_{L^\prime}^2(k_{ik}y)
\sum_{M_LM_{L^\prime}}
|Y_{LM_{L}}(\hat{k_{ij}})|^2|Y_{L^\prime M_{L^\prime}}(\hat{k_{ik}})|^2
\hspace{.5cm}\nonumber\\
\sum_{\{\sigma_i \sigma_j \sigma_k\}}
\sum_{M_SM_{S^\prime}}
{\left(\matrix{
{1\over 2}&{1\over 2}&S\cr
\sigma_i & \sigma_j&M_{S}
\cr}\right)}^2
{\left(\matrix{
{1\over 2}&{1\over 2}&{S^\prime}\cr
\sigma_i & \sigma_k&M_{S^\prime}
\cr}\right)}^2
\hspace{5.5cm}\nonumber\\
\sum_{\{\tau_i \tau_j \tau_k\}}\sum_{M_TM_{T^\prime}}
{\left(\matrix{
{1\over 2}&{1\over 2}&T\cr
\tau_i & \tau_j&M_{T}
\cr}\right)}^2
{\left(\matrix{
{1\over 2}&{1\over 2}&{T^\prime}\cr
\tau_i & \tau_k&M_{T^\prime}
\cr}\right)}^2\hspace{5.5cm}\nonumber\\
\sum_{M_JM_{J^\prime}}
{\left(\matrix{
L&S&J\cr
M_L&M_{S}&M_{J}
\cr}\right)}^2
{\left(\matrix{
L^\prime&S^\prime&J^\prime\cr
M_{L^\prime}&M_{S^\prime}&M_{J^\prime}
\cr}\right)}^2\hspace{5.5cm}
\end{eqnarray}
where the large parentheses are the familiar Clebsch-Gordan coefficients and

\begin{center}
$$
h^{LSJT}(x)=<LSJT\mid h(x) \mid LSJT> \hspace{1cm} {\cal V}%
^{LSJT}(y)=<LSJT\mid {\cal V}(y)\mid LSJT>
$$
$$
\vec x=\vec r_1-\vec r_3 \hspace{1.75cm} \vec y=\vec r_1-\vec r_2
\hspace{1.75cm} \vec k_{ij}={\frac{1 }{2}}(\vec k_i-\vec k_j)
$$
$$
\vec k_{ik}={\frac{1 }{2}}(\vec k_i-\vec k_k) \hspace{1.75cm} \vec
K_{ik}=\vec k_i+\vec k_k \hspace{1.75cm} \vec K_{ij}=\vec k_i+\vec k_j
$$
\end{center}

$\sigma$'s and $\tau$'s stand for particles spin and isospin respectively.
The following identity,
\begin{eqnarray}
\sum_{l_1l_2m_1m_2}j_{l_1}(qx)j_{l_1}(px)j_{l_2}(qy)j_{l_2}(py)
Y_{l_1m_1}(\hat{q})Y_{l_1m_1}^*(\hat{p})
Y_{l_2m_2}(\hat{p})Y_{l_2m_2}^*(\hat{q})=\hspace{3cm}\nonumber\\
{1 \over {(4\pi)^4}} \int d\Omega_xd\Omega_y \exp{(i(\vec{x}-\vec{y})\cdot
(\vec{q}-\vec{p}))}\hspace{1cm}
\end{eqnarray}
is used to check the validity of equation (34). To see this matter clearly
we assume that ${\cal V}$ and $h$ are state-independent, then, it is
possible to perform all of the summations and we show that one can reach to
our previous equations for the state-independent three-body cluster energy
(equation (30-33)).

In figures 3 and 4 we plot $E_{3h}^{(2)}$ by using the state-dependent
correlation functions and effective potentials i.e. $f_\alpha ^{(i)}(r)$ and
${\cal V}_\alpha ^{(i)}(r)$ for Reid and $\Delta $-Reid Potentials
respectively. $E_{3h}^{(2)}(C)$ shows the same calculations but only with
the central parts of $f_\alpha ^{(i)}(r)$. $E_{3h}^{(2)}(NC)$ gives the
estimate of non-central parts i.e. $E_{3h}^{(2)}$-$E_{3h}^{(2)}(C)$. From
these figures, as one should expect, it can be concluded that the main
contribution in $E_{3h}^{(2)}$ comes from the central parts of correlation
functions and with very good approximation we can ignore the non-central
parts of energy contributions in the three-body expressions (this is not a
good approximation in the two-body level).

The exact calculation of $E_{3h}^{(3)}$ is very complicated, but according
to the above argument we can take into the account only the central part of
correlation functions and make it possible to have an estimate for this term
as well i.e.
\begin{equation}
E_{3h}^{(3)}(c)=-{\frac 1{{A}}}\sum_{ijk}<ijk\mid h({13}){\cal {V}}({12}%
)\mid ijk>_a
\end{equation}
We start by inserting a unit operator $\sum_{lmn}\mid lmn><lmn\mid $ between
the two-body operators $f(ij)$ and ${\cal V}(ij)$ in above equation.
\begin{equation}
E_{3h}^{(3)}(c)={\frac 1A}\sum_{ijk,lmn}<ijk\mid h(13)\mid lmn><lmn\mid
{\cal V}(12)\mid ijk>_a
\end{equation}
Now by expanding the anti-symmetrized ket we get 9 terms which are very
similar when we want to calculate them. We explain how it is possible to
calculate one of these terms i.e.
\begin{eqnarray}
\sum_{ijklmn}<ijk \mid h(13)\mid lmn><lmn \mid {\cal V}(12) \mid ikj-kij>=
\hspace{5cm}\nonumber\\
\sum_{ijkl}<ik \mid h(13)\mid lj><lj \mid {\cal V}(12) \mid ik-ki>=
\sum_{\{k_ik_jk_kk_l\}}\sum_{LSJT}\sum_{L^\prime J^\prime}
\int_0^\infty\int_0^\infty \, dxdy \,
\hspace{1cm}\nonumber\\
(xy)^2[T]({{(4\pi)^4(2\pi)^3}\over
{\Omega^3}})((-1)^{L+S+T}-1)
h^{LSJT}(x){\cal V}^{L^\prime SJ^\prime T}(y)\hspace{3cm}\\
J_L(k_{ik}x)J_L(k_{lj}x)J_{L^\prime}(k_{ik}y)J_{L^\prime}(k_{lj}y)
\hspace{4.3cm}\nonumber\\\sum_{M_LM_{L^\prime}}
Y_{LM_L}(\hat{k_{ik}})Y_{LM_L}^*(\hat{k_{lj}})
Y_{L^\prime M_{L^\prime}}(\hat{k_{lj}})
Y_{L^\prime M_{L^\prime}}^*(\hat{k_{ik}})\hspace{4cm}\nonumber\\
\sum_{M_JM_{\prime{J}}M_S}
{\left(\matrix{
L&S&J\cr
M_L&M_{S}&M_{J}
\cr}\right)}^2
{\left(\matrix{
{L^\prime}&S&J^{\prime}\cr
M_{L^\prime}&M_{S}&M_{J^{\prime}}
\cr}\right)}^2\delta((\vec k_i+\vec k_k)-(\vec k_l+\vec k_j))
\nonumber
\end{eqnarray}
In figures 5 and 6 we plot the various quantities such as $E_{3h}^{(2)}(C)$,
$E_{3h}^{(3)}(C)$ and $E_{3h}(C)$=$E_{3h}^{(2)}(C)$ +$E_{3h}^{(3)}(C)$ for
the Reid and the $\Delta $-Reid potentials respectively. $\bar
E_{3h}^{(2)}(C)$, $\bar E_{3h}^{(3)}(C)$ and $\bar E_{3h}(C)$ are the same
terms but with the state-averaged approximation of section III-A.

Finally in figures 7 and 8 we plot the total contribution of $E_{3h}(C)$ by
presenting the exact and the state-averaged calculation for the Reid and the
$\Delta$-Reid interactions. The size of $E_{3h}(C)$ is less than
state-averaged $\bar E_{3h}(C)$ and this difference becomes sizable beyond
the nuclear matter saturation density. This shows that at large densities
the state-averaged approximation is no longer valid. Our final results for
the binding energy of nuclear matter per nucleon are given in table 2 by
adding the three-body cluster energies to our LOCV calculations.

\section{Discussion}

We have developed a method to make a very good estimate of the three-body
cluster energy by using the state-dependent hole functions and effective
interactions. We found that the size of the three-body cluster energy is
much more smaller than our previous estimate using the state-averaged
approximation. The smallness of the three-body cluster energy indicates that
we are very close to the exact results at LOCV level and our cluster
expansion series converges very rapidly. Our result with the $\Delta$-Reid
interaction is comparable with those calculations in which the three-body
interactions [5-9] have been taken into the account such that they could
reproduce the empirical properties of nuclear matter [5-9]. Presumably in
the lowest order limit the cluster expansion for the exact two-body radial
distribution function, $g(r)$, is automatically convergent and, provided
that the lowest order energy has been efficiently minimized, the replacement
of $g(r)$ by its lowest order approximation,
\begin{equation}
g^{(2)}(r)=\{1-[{\frac{3 }{{2}}}l(k_Fr]^2\}f^2(r)
\end{equation}
is then a very good approximation. In this case the integral constraint
(normalization constraint),
\begin{equation}
\rho \int \,dr_{12}\, [1-g(r_{12}]=1
\end{equation}
plays a crucial role at all densities in forcing the correlation function to
heal and in restricting the size of the wound in the correlated two-body
wave function.

It is interesting to note that it may also be the inclusion of the tensor
and spin-orbit forces and the incorporation of tensor correlation which
allows  us to deal accurately with the contributions to the energy of
nuclear  matter from all of the states of realistic potentials and to obtain
such  good agreement with far more ambitions and complicated calculations
such as VHC.

Whether or not the above arguments are quantitatively correct, we still have
to face that for $\rho \leq 0.3fm^{-3}$ our nuclear matter results agree
extremely well with those of VHC, which are hopefully reliable upper bounds
on the energy up to this density and differ considerably from the accepted
experimental equilibrium values. So we can conclude that only by taking into
accounts the internal degrees of freedom of nucleons (such as $\Delta (1234)$%
) [2], the three-body forces [5-9] or the  relativistic effect [16] it may
be possible to remove this discrepancy between the theoretical calculation
and the empirical prediction.

In this context we remark again that the LOCV method because of (1) its
agreement with VHC which includes the many-body cluster contributions, (2)
the smallness of three-body cluster energy and (3) its great simplicity, is
a useful tool in the study of other properties of nuclear matter and finite
nuclei.

\newpage

\newpage

\ \\ {\bf Figure captions :} \\ {Figure 1: Various terms of the three-body
cluster energy (MeV) versus density ($fm^{-3}$) with the state-averaged
correlation function and the effective interaction in nuclear matter
according to the equations of section III-A for the Reid interaction. \\} \\
{Figure 2: Same as figure 1 but for the $\Delta $-Reid interaction.\\} \\ {%
Figure 3: The plot of $E_{3h}^{(2)}$ (MeV) versus density ($fm^{-3}$) by
doing an exact calculations with the Reid potential. $E_{3h}^{(2)}(C)$ and $%
E_{3h}^{(2)}(NC)$ are the central and non-central parts of this term,
respectively. \\} \\ {Figure 4: Same as figure 3 but for the $\Delta $-Reid
potential.\\} \\ {Figure 5: Comparisons of $E_{3h}(C)$ (MeV), etc. with $%
\bar E_{3h}(C)$ etc. versus density ($fm^{-3}$) for the Reid potential\\} \\
{Figure 6: Same as figure 5 but for the $\Delta $-Reid potential.\\} \\ {%
Figure 7: The plots of the state-averaged $\bar E_{3h}$ and its exact
central contribution, $E_{3h}(C)$ (MeV) versus density ($fm^{-3}$) for the
Reid interaction. $\bar E_{3h}(C)$ is the central part of state-averaged
contribution \\ {Figure 8: Same as figure 7 but for the $\Delta $-Reid
potential.\\} \newpage
\ {\bf Table 1.} The saturation energy, the saturation density and the
incompressibility of nuclear matter for different potentials and the
many-body methods (see different abbreviation and references on page 2). }

\begin{center}
\begin{tabular}{cccccc}\hline\hline
Potential       &  Method   & Author        &$\rho_0 (fm^{-3})$&E(MeV)  & ${\cal K}$(MeV)\\ \hline
$AV_{18}$       &  LOCV     &   BM[6]       &     0.310        &     -18.46&302           \\
$AV_{14}$       &  LOCV     &   BM[6]       &     0.290        &   -15.99  &248         \\
                &VHC        &  WFF[8]       &     0.319        &   -15.60  &205
\\
                &   BB      &   DW[7]       &     0.280        &   -17.80  &247
\\
                &  BHF      &  BBB[9]      &     0.256        &   -18.26  &-
\\
$UV_{14}$       &  LOCV     &   BM[5]       &     0.366        &   -21.20  &311         \\
                &VHC        &   CP[8]       &     0.349        &   -20.00  &-
\\
                &VHC        &  WFF[8]       &     0.326        &   -17.10  &243
\\
$UV_{14}$+TNI   &   LOCV    &   BM[5]       &     0.170        &   -17.33  &276         \\
                &VHC        &  WFF[8]       &     0.157        &   -16.60  &261
\\
                &  CBF      &  FFP[8]       &     0.163        &   -18.30  &269
\\
$\Delta$-Reid   &  LOCV     &   MI[2]       &     0.258        &   -16.28  &300         \\
Reid            &  LOCV     &  OBI[1]       &     0.294        &   -22.83  &340         \\
                &  LOCV     &  MO[8]       &     0.230        &   -14.58  &238
\\
Empirical       &    -      &   -        &     0.170         &   -15.86    &(200-300)       \\  \hline\hline
\end{tabular}
\end{center}

\ {\bf Table 2.} The saturation energy and the saturation density of nuclear
matter by adding the three-body cluster energy according to the text.

\begin{center}
\begin{tabular}{cccc|ccc}\hline
                 &\multicolumn{3}{c|}{Reid} &\multicolumn{3}{c}{$\Delta$-Reid}\\ \hline\hline
                 &  $E_2$    &$E_2+\bar E_3$&  $E_2+E_3$    &  $E_2$   & $E_2+\bar E_3$& $E_2+E_3$    \\ \hline
$\rho_0(fm^{-3})$&   0.294   &     0.216    &       0.295   &   0.258  &     0.208     &      0.257   \\
E(MeV)           & -22.83    &   -22.24     &       -23.362 &   -16.74 &     -15.597   &      -16.721 \\ \hline
\end{tabular}
\end{center}

\end{document}